\documentclass[apjl]{emulateapj}
\usepackage{graphicx,color} \usepackage{amsmath} \usepackage{times}
\usepackage{hyperref}
\hypersetup{
    colorlinks,
    linkcolor={blue},
    citecolor={blue},
    urlcolor={blue}
}
\def\apj{Astrophys.\ J.}
\def\apss{Astrophys.\ Space Sci.}
\def\mnras{Mon.\ Not.\ R.\ Astron.\ Soc.}
\def\aap{Astron.\ Astrophys.}
\def\apjl{Astrophys.\ J.\ Lett.}

\def\jfm{J.\ Fluid Mech.}
\def\jcap{J.\ Cosmol.\ Astropart.\ Phys.}
\def\physrep{Phys.\ Rep.}

\def\prl{Phys.\ Rev.\ Lett.}
\def\prd{Phys.\ Rev.\ D}

\def\ap{Astropart.\ Phys.}

\def\ssr{SSRv}

\def\araa{Ann.\ Rev.\ Astron.\ Astrophys.}
\def\jtb{J.\ Theor.\ Biol.}

\def\jap{J.\ Appl.\ Probab.}

\newcommand{\CR}{cosmic ray}

\newcommand{\Lar}{r_{\text{L}}}    

\newcommand{\Rm}{R_\mathrm{m}} 
\newcommand{\Rmc}{R_\mathrm{m, c}}    

\renewcommand{\vec}[1]{\mathbf{#1}}	
\newcommand{\dd}{\mathrm{d}}        
\newcommand{\ii}{\mathrm{i}}        
\newcommand{\RL}{\Lar}
\newcommand\deriv[2]{\displaystyle\frac{\partial #1}{\partial #2} }
 \newcommand\Fig[1]{Fig.~\ref{#1}}
\newcommand{\cm}{\,{\rm cm}}    
\newcommand{\p}{\,{\rm pc}}     
\newcommand{\kpc}{\,{\rm kpc}}  

\newcommand{\mkG}{\,\mu{\rm G}} 

\newcommand{\GeV}{\,{\rm GeV}}  

\begin{document} 
\title{Cosmic Rays in Intermittent Magnetic Fields}
\author{Anvar \surname{Shukurov}}\affiliation{School of Mathematics and
Statistics, Newcastle University, Newcastle Upon Tyne, NE1 7RU, UK}
\author{Andrew P. \surname{Snodin}}\affiliation{Department of Mathematics, Faculty
of Applied Science, King Mongkut's University of Technology North Bangkok,
Bangkok 10800, Thailand}\author{Amit \surname{Seta}}\thanks{a.seta1@ncl.ac.uk;
amitseta90@gmail.com}\author{Paul J.
\surname{Bushby}}\author{Toby S. \surname{Wood}}
\affiliation{School of Mathematics and Statistics, Newcastle University,
Newcastle Upon Tyne, NE1 7RU, UK}

\begin{abstract} 
The propagation of cosmic rays in turbulent magnetic fields is a diffusive
process driven by the scattering of the charged particles by random magnetic
fluctuations. Such fields are usually highly intermittent, consisting of
intense magnetic filaments and ribbons surrounded by weaker, unstructured
fluctuations. Studies of cosmic ray propagation have largely overlooked
intermittency, instead adopting Gaussian random magnetic fields. Using test
particle simulations, we calculate cosmic ray diffusivity in intermittent,
dynamo-generated magnetic fields. The results are compared with those obtained
from non-intermittent magnetic fields having identical power spectra. The
presence of magnetic intermittency significantly enhances cosmic ray diffusion
over a wide range of particle energies. We demonstrate that the results can be
interpreted in terms of a correlated random walk.
\end{abstract}
\maketitle
\section*{Introduction} Cosmic rays are charged relativistic particles (mostly
protons) scattered, as they propagate, by random
magnetic fields \citep{G90}. Over sufficiently long time and length scales, their
propagation is diffusive \citep{Cesarsky1980}. Assuming an  interstellar magnetic field of strength $5 \mkG$, the Larmor 
radius $\Lar$ of a \CR\ proton of energy $5 \GeV$ is of order
$10^{12} \cm$, much smaller than the correlation length of
interstellar MHD turbulence ($\sim10^{20}\cm$).
Thus, \CR s closely follow field lines (for a significant time) and so the
geometry and statistical properties of magnetic fields control their
propagation.  The dominant contribution to particle scattering is from magnetic
irregularities at a scale comparable to $\Lar$. In this paper,
we mostly discuss cosmic rays that propagate diffusively.

With exceptions discussed below \citep[see also][]{AB_lR2014,PucciEA2016}, studies of \CR\
propagation employ random magnetic fields with Gaussian statistics that are
completely described by the two-point correlation function or the power spectrum
\citep[e.g.,][]{MichalekOstrowski97,GJ1999,Casse_et_al2002,Parizot2004,
CandiaRoulet2004,DBS2007,GAP08,Plotnikov_et_al2011,HMR14,Snodin_et_al2016,Subedi2016}.
However, the interstellar and intergalactic magnetic fields have a more
complicated structure. The fluctuation (small-scale) dynamo
\citep{ZRS90, Wilkin_et_al2007} and random shock waves \citep{ByT87} produce
highly intermittent, strongly non-Gaussian,
essentially three-dimensional magnetic fields with random magnetic
filaments and ribbons surrounded by weaker fluctuations. 
Filamentary and planar structures in the interstellar medium, consistent 
with the notion of spatial intermittency, 
have been detected in the radio \citep[Sect.~5.2 in][]{HS13}
and sub-millimeter \citep{Z+15} ranges as well as in the neutral hydrogen 
distribution \citep{HT15}.
In such a magnetic
field, the propagation of charged particles is controlled not only by its power
spectrum, but also by the size and separation of the magnetic structures.  The
influence of such a complex magnetic field upon \CR\ propagation is poorly
understood. Existing theories, on the quasilinear approach
\citep[][]{Jokipii1966,Sch02,G90}, or its nonlinear extensions and
alternative ideas \citep[e.g.,][]{MatthaeusEA2003,Sh09,YanLazarian2002,VSMB98},
do not consider intermittency, or use the Corrsin hypothesis \citep[][]{Corrsin59}, which
assumes Gaussian statistics for the magnetic field.
Recent test particle simulations used magnetic fields obtained from
simulations of MHD turbulence \citep[e.g.,][]{DmitrukEA2004,RevilleEA2008,BYL2011,LynnEA2012,
WeidlEA2015,CohetMarcowith2016} \citep[see also][]{RohEA2016}.
These models are free from the assumption of Gaussian statistics but they do not
consider any effects of magnetic structures even if those were present.
There have been no systematic attempts  
to examine the significance of realistic, physically realizable magnetic intermittency
in 3D;
this is our goal here. In intermittent magnetic fields, particle trapping can be important even in 3D.
We note that the Kubo number, often used to delineate different transport regimes, 
depends only on second-order correlations and is therefore insensitive to intermittency.

We use test particle simulations \citep{GJ1999,Casse_et_al2002, DZ2014, Snodin_et_al2016},
integrating the equation of motion for 
a large number of particles in a 
statistically isotropic, prescribed magnetic field,
in the regime where cosmic ray pressure is too low to excite significant MHD waves.
The magnetic field is obtained as a solution of the induction equation with a prescribed velocity field
that drives the fluctuation dynamo.  This produces a realistic, intermittent
magnetic field. The degree of intermittency depends on the magnetic Reynolds
number $\Rm$. As $\Rm$ increases,
the magnetic structures occupy a smaller proportion of the volume.  The
intermittency introduces two distinct particle propagation regimes, one within a
magnetic structure and another between them. Cosmic ray particles are
strongly scattered by the magnetic structures and move relatively freely between
them. By comparing particle diffusion in an intermittent field with that in a
magnetic field lacking structure, but with identical power spectrum, we
demonstrate that intermittency can significantly enhance diffusion, and so
diffusion cannot be described in terms of the power spectrum alone.
\section*{Magnetic Field Produced by Dynamo Action} We generate
intermittent, statistically isotropic, fully three-dimensional random magnetic
fields $\vec{b}$ 
by solving the induction equation with a
prescribed velocity field $\vec{u}$, 
\begin{equation}\label{inductioneqn}
\deriv{\vec{b}}{t} = \nabla \times (\vec{u} \times \vec{b}) 
+ \Rm^{-1} \nabla^2 \vec{b}, 
\qquad
\nabla\cdot \vec{b}=0, 
\end{equation} 
with
periodic boundary
conditions in a cubic domain of width $L=2\pi$
and
$256^3$ or $512^3$ mesh points. 
Equation~\eqref{inductioneqn} is written in a dimensionless
form, expressing length in the units of the flow scale $l_0$ and time in the
units of $l_0/u_0$, where $u_0$ is the rms flow speed.  Here $\Rm=l_{0}u_0/\eta$
is the magnetic Reynolds number \footnote{Some authors define $\Rm$ in
terms of the wavenumber $k_0$, resulting in $\Rm$ values a factor of $2\pi$
smaller.} and $\eta$ is the magnetic diffusivity, assumed to be constant. 
In a generic, three-dimensional, random
flow, dynamo action occurs 
(i.e., the mean magnetic energy density 
grows exponentially with $t$) 
provided 
$\Rm>\Rmc$, where
$\Rmc$ is the critical magnetic Reynolds number \citep{ZRS90}.  Depending on the nature of the
velocity field, typically $\Rmc\simeq10$--$100$, and the magnetic field decays
for $\Rm<\Rmc$  \citep{BS2005}.
As $\Rm\to\infty$, the magnetic structures produced by the dynamo become
progressively more filamentary in  nature, with the thickness of each filament
of the order of $d=l_0\Rm^{-1/2}$, and a characteristic filament length (radius
of curvature) of the order of $l_0$ \citep{ZRS90,Wilkin_et_al2007}. 
The magnetic field used in our simulations is an eigenfunction
obtained by renormalizing the exponentially growing solution of
Eq.~\eqref{inductioneqn} to have a constant rms field strength $b_0$.
We expect the magnetic
structure of the corresponding nonlinear dynamo to be similar to that of the
marginal eigenfunction
obtained at $\Rm\approx\Rmc$ \citep{Sub99}.
However, we consider a wider range of $\Rm$ to explore the effects of
a variable degree of intermittency: it increases with $\Rm$.
\begin{figure*} \centering
      \includegraphics[width=2\columnwidth]{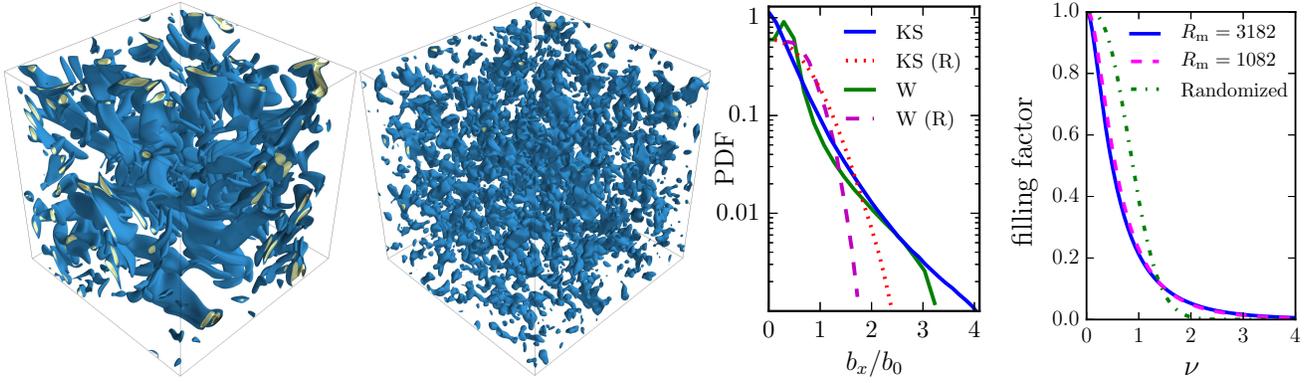} \caption{Isosurfaces of
	magnetic field strength $b^2/b_0^2 = 2.5$ (blue) and $b^2/b_0^2 = 5$ (yellow)
	with $b_0$ the rms magnetic field, for magnetic field generated by the KS flow
\eqref{KS} at $\Rm = 1082$ (left) and for the same magnetic field after Fourier
phase randomization as described in the text (second from left). Magnetic field
generated by the W flow \eqref{willis_vel} is similarly affected (not shown). The
	second from right panel shows the PDFs of a magnetic field component $b_{x}$ for the
original (KS,W: solid) and randomized [KS (R), W (R): dashed] magnetic fields
  obtained with both velocity fields (only $b_{x}>0$ is show as the PDFs are
  essentially symmetric about $b_x=0$). The randomized fields have almost perfectly
Gaussian statistics, whereas magnetic intermittency leads to heavy tails.
The panel on the right  shows the fractional volume within magnetic structures where $b\geq \nu b_0$, with $b_0$ the rms
      field strength, as a function of $\nu$ for the intermittent magnetic field 
      produced by the flow \eqref{KS} (solid for $\Rm=3182$ and dashed for $\Rm=1082$) and its Gaussian counterpart 
      (dash-dotted, for $\Rm=3142$ and $1082$)
      obtained by Fourier phase randomization; 
      the filling factor of the randomized fields is independent of $\Rm$.
      \label{structure}} \end{figure*}
To isolate robust features of cosmic ray propagation independent
of the particular form of intermittent magnetic field, we
use two types of incompressible flow to drive the dynamo, both chaotic, but
one of a single scale, and the other multi-scale with a controlled power
spectrum.  The first flow \citep{Willis2012}, henceforth referred to as flow W,
is stationary, 
\begin{equation}\label{willis_vel} 
\vec{u}(\vec{x}) =
(2/\sqrt3)(\sin y \cos z, \sin z \cos x, \sin x \cos y).  
\end{equation} 
It is a
very efficient dynamo with $\Rmc\approx 11$, producing regularly spaced magnetic
structures
in
the form of ellipsoids of identical
size that become thinner as $\Rm$ increases and whose positions are determined
solely
by the
flow geometry (so are independent of $\Rm$).  The
second flow (KS) is time-dependent and multi-scale; it was employed for dynamo
simulations \citep{Wilkin_et_al2007} and as a Lagrangian model of turbulence
\citep{FungEA1992}: 
\begin{equation} \label{KS} 
\vec{u}(\vec{x},t) =
\sum_{n=0}^{N-1} \left(\vec{C}_{n} \cos \phi_n + \vec{D}_{n} \sin \phi_n\right),
\end{equation} 
where $\phi_n = \vec{k}_n \cdot \vec{x} + \omega_n t$, with
$\vec{k}_n$ a randomly oriented wave vector (of magnitude $k_n$) and $\omega_n$
a 
frequency specified below. The random vectors, $\vec{C}_n$ and $\vec{D}_n$, are
chosen to be orthogonal to $\vec{k}_n$ to ensure $\nabla \cdot \vec{u} = 0$. We
select $N=40$ distinct wave vectors, with magnitudes between $k_0 = 2 \pi / L$
and $k_{N-1} \approx 8 k_0$, so that the flow is periodic with the outer scale
$l_0 = L$.  The amplitudes of $\vec{C}_n$ and $\vec{D}_n$ are selected to
produce an energy spectrum $E(k) \propto k^{-5/3}$ with $\int_{k_0}^{k_{N-1}}
E(k)dk = u_0^2/2$. We take $\omega_n=[k_n^3 E(k_n)]^{1/2}$, which introduces 
a scale-dependent time variation.  The dynamo in this flow has $\Rmc \approx
1000$ \citep{Wilkin_et_al2007}. The flow produces transient magnetic structures,
consisting of filaments of various sizes, as illustrated in the leftmost panel
of \Fig{structure}.

To 
identify
the effect of magnetic intermittency on \CR\ diffusion, we
also consider random magnetic fields where the structures have been destroyed
but the magnetic energy spectrum remains unchanged \citep{Snodin_et_al2013}.
This is achieved by taking the spatial Fourier transform of $\vec{b}(\vec{x})$
from Eq.~\eqref{inductioneqn}, and then multiplying each complex Fourier mode by
$\exp[\ii \psi(\vec{k})]$, with $\psi(\vec{k})$ a random phase selected
independently for each $\vec{k}$. The inverse Fourier transform of the result
produces a magnetic field with an unchanged spectrum but with little remaining
structure, as demonstrated in the second from left panel
of \Fig{structure}.  As shown on
the second from right panel of \Fig{structure}, the probability density functions (PDFs)
of the field components for the intermittent fields produced by each flow (W and
KS) have long, heavy tails, while the phase randomization produces nearly
Gaussian random fields.
Another aspect of this difference is also illustrated in the rightmost panel of
Fig.~\ref{structure} where the fractional volume 
occupied by magnetic structures with $b/b_0>\nu$ is shown as a function of $\nu$: 
an intermittent magnetic field has more strong, localized structures with $\nu\gtrsim1.4$ than
a Gaussian field with identical power spectrum.

To explore the effects of a mean magnetic field,
we also consider particle propagation in a magnetic field given by 
$\vec{B}=\vec{b}+\vec{B}_0$, where $\vec{B}_0$ is an imposed uniform magnetic
field. In such cases, the rms magnetic field $\tilde{b}_0$ quoted below includes the 
mean part,
$\tilde{b}_0^2=B_0^2+b_0^2$.
\section*{Cosmic Ray Propagation} Using magnetic field realizations generated
from Eq.~\eqref{inductioneqn}, or the corresponding randomized magnetic fields,
we obtain an ensemble of \CR\ trajectories ($\ge 1000$ in number) by solving
numerically the 
dimensionless equation of motion 
for the particle trajectories $\vec{x}(t)$,
\begin{equation}  \label{eq:Newton} 
\ddot{\vec{x}} = \alpha\dot{\vec{x}}\times\vec{B}(\vec{x}),
\end{equation}
with
$\alpha = ql_0\tilde{b}_0/(\gamma m c v_0)$, $q$ the particle
charge, $m$ its rest mass, $\tilde{b}_0$ the total rms field strength,$\gamma$ 
the Lorentz factor, $v_0$ the particle speed and $c$ the speed of light. 
As in most cosmic ray propagation models \citep{G90,Sch02,Sh09}, we  
neglect electric fields in Eq.~\eqref{eq:Newton}:
they are negligible at the scales of interest 
($\simeq1\kpc$ in galaxies and $\simeq10\kpc$ in galaxy clusters).
Hence, the particle speed $v_0$ remains constant.
Each particle is given a random initial position and propagation direction, 
but the same initial speed.
The {\it characteristic} dimensionless Larmor radius,
based on the rms magnetic field strength,
is $\Lar/l_0 = \alpha^{-1}$;
we use this ratio to characterize the particle properties.
When $B_0=0$, we
calculate the isotropic
diffusion coefficient $\kappa= \lim_{t \to \infty} \langle |\Delta\vec{x}(t)|^2
\rangle / (6t)$, where $\Delta\vec{x}(t)$ is the particle displacement, and the
angular brackets denote averaging over particle displacements.
In the presence of a mean magnetic field directed along the $z$-axis,
we introduce similarly defined parallel and perpendicular diffusion coefficients, 
$\kappa_\parallel=\lim_{t \to \infty} \langle \Delta z(t)^2\rangle /(2t)$
and $\kappa_\perp=\lim_{t \to \infty} \langle [\Delta x(t)^2+\Delta y(t)^2]\rangle /(4t)$.
\begin{figure} \centering \includegraphics[width=\columnwidth]{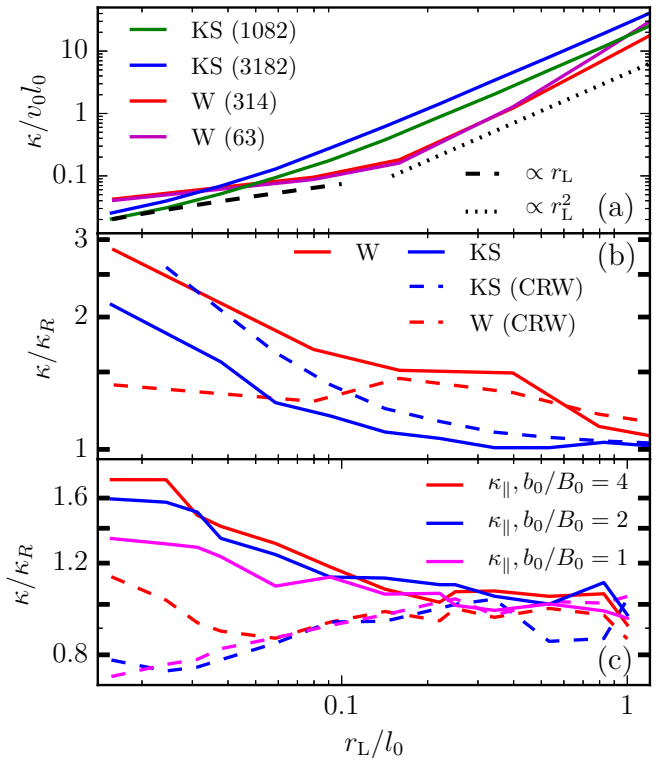}
\caption{\textbf{(a)} The \CR\ diffusion coefficient for the W flow
	\eqref{willis_vel} (red, magenta) and the KS flow \eqref{KS}
            (blue, green) as a function of $\RL/l_0$ for the values of $\Rm$ given
in parenthesis in the legend.  The dotted and dashed line shows the scaling $\kappa\propto
\Lar^2$ and $\kappa \propto \Lar$ respectively.  \textbf{(b)} The ratio of diffusion coefficients from intermittent,
$\kappa$,  and randomized, $\kappa_\mathrm{R}$, magnetic fields for the two
	flows (solid lines, KS with $\Rm=3182$, W with $\Rm=314$). The dashed lines of same color show the corresponding CRW model, Eq.~\eqref{dcrw}.
\textbf{(c)} As in (b) but in the presence of a mean magnetic field $B_0$,
of the relative strength specified in the legend, for the KS flow with $\Rm=3182$; 
$\kappa_\parallel$ and $\kappa_\perp$ are shown solid and dashed, respectively.}
\label{dw} 
\end{figure}
\begin{figure*} \centering \includegraphics[width=2\columnwidth]{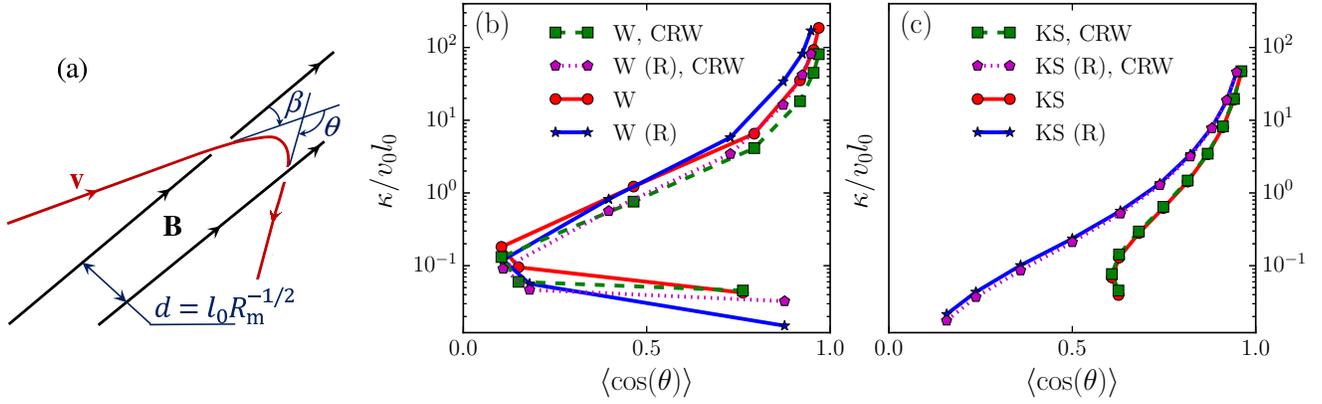}
      \caption{\textbf{(a)} A charged particle with the pitch angle $\beta$ is
deflected by an angle $\theta$ in a magnetic structure of a thickness $d$, thus
	introducing correlation into the random particle trajectory.
	The dependence of the diffusivity in simulations on
	$\langle\cos\theta\rangle$ in the W flow and the randomized (R) magnetic field is shown in \textbf{(b)} (solid lines). The corresponding CRW approximations (Eq.~\eqref{dcrw}) are shown with broken lines.
\textbf{(c)} is as in (b), but for the KS flow.
These results are obtained for $B_0=0$.} 
\label{crw} 
\end{figure*}
\section*{Cosmic Ray Diffusivity}
Figure~\ref{dw}a shows the dependence of the \CR\ diffusion coefficient on
$\RL/l_0$ (proportional to the particle energy)
for $B_0=0$. 
For $\RL/l_0\gg1$,
we recover the asymptotic scaling $\kappa \propto \RL^2$ (high energy limit) in agreement with earlier
results 
{\citep{Parker65,AloisioBerezinsky2004,Parizot2004,GAP08,DBS2007,BYL2011,
Plotnikov_et_al2011,HMR14,Snodin_et_al2016,Subedi2016}.
At lower energies, the dependence of $\kappa$ on particle energy is weaker and is sensitive to magnetic structure.
Magnetic intermittency is expected to be important at those energies where
\begin{equation}\label{range} 
\Lar/l_0\lesssim1,  
\end{equation} and the
dependence $\kappa(\Lar/l_0)$ 
in Fig.~\ref{dw}a
indeed deviates from the asymptotic form in this range.
The role of magnetic intermittency is demonstrated in \Fig{dw}b,
showing the
ratio of the diffusivity $\kappa$ calculated with a dynamo generated 
magnetic field to
that in the corresponding randomized field, $\kappa_\mathrm{R}$
($B_0=0$ in Panels a and b).  At high
energies (large 
$\Lar/l_0$}), $\kappa/\kappa_\mathrm{R}\simeq1$, suggesting that the
magnetic structures play little role.  However, $\kappa/\kappa_\mathrm{R}$
increases rapidly up to more than $2.5$ at lower energies: magnetic structures
enhance diffusion when inequality~\eqref{range} is satisfied. 
We find that the ratio
$\kappa/\kappa_\mathrm{R}$ 
at fixed
$\RL/l_0$ 
increases
with
$\Rm$ for a given flow.  At high values of 
$\RL/l_0$, 
the diffusivity
still depends on $\Rm$ via changes in the magnetic correlation length (\Fig{dw}a), but not via the $\Rm$-dependent intermittency,
as suggested by \Fig{dw}b where $\kappa/\kappa_R$ tends to unity as $\Lar/l_0$ increases.
One might expect a change in the diffusivity behavior at $\RL/l_0 \approx \Rm^{-1/2}$, associated
with the 
thickness of magnetic filaments,
and this may explain the variation in slope of
$\kappa$ at low $\RL/l_0$ in \Fig{dw}a (or the ratios in \Fig{dw}b). However, at
present the role of this scale is unclear.

Figure~\ref{dw}c 
illustrates the effects of the mean magnetic field, presenting the ratio
of the parallel and perpendicular diffusivities in the intermittent and Gaussian magnetic fields.
A mean magnetic field somewhat reduces the effect of intermittency, but does not eliminate it
even for $b_0/B_0=1$. Magnetic intermittency
enhances $\kappa_\parallel$ (i.e. $\kappa_\parallel > (\kappa_\parallel)_\mathrm{R}$) at all but the highest energies, but $\kappa_\perp<(\kappa_\perp)_\mathrm{R}$ at lower energies for $b_0/B_0=2$ and $b_0/B_0=1$. The effects of the mean field will be discussed in detail elsewhere.
\section*{Cosmic Ray Propagation as a Correlated Random Walk} 
The Brownian motion is a widely used model for diffusive processes. 
This is the simplest type of random walk where each step is made in a direction independent of 
the previous direction. In a continuum limit, it leads to the diffusion equation. However, a charged particle
moves differently. As illustrated in Fig.~\ref{crw}a, the direction of its
motion after deflection by a magnetic structure is correlated with the previous
direction. The deflection angle $\theta$ is related to $\Lar$, the angle
between the velocity and magnetic field, $\beta$, and the magnetic
structure width $d$, 
\begin{equation}\label{theta} 
\theta\simeq d/(\Lar\sin\beta).  
\end{equation} 
This is a \textit{correlated random walk}
(CRW) \citep{G55}, a first-order Markov chain (since the correlation does not
extend beyond two consecutive steps).  For a symmetric probability distribution of
$\theta$, the CRW diffusivity depends on
$\langle\cos\theta\rangle$, where angular brackets denote the ensemble average.
The mean-square displacement in the CRW was obtained in 2D \citep{KS1983}, and implies
the following 3D diffusivity \citep[Eq.~(3.3.7) in][]{CR92}:
\begin{equation}\label{dcrw} 
\kappa = \frac{\langle l^2 \rangle}{6\tau} +
\frac{\langle l \rangle ^2}{3\tau} \frac{\langle\cos\theta\rangle}{1 -
\langle\cos\theta\rangle}, 
\end{equation} 
with $\tau=\langle l\rangle/v$, $v$
the particle speed and $l$ the step length.  To calculate $\langle\cos\theta\rangle$, we
assume that the pitch angle $\beta$ is uniformly distributed between $0$
and $\pi$. Defining $a=d/\Lar\simeq l_0\Rm^{-1/2}/\Lar$, it can be shown that
\begin{align}\label{ct}
\langle\cos\theta\rangle&=\pi^{-1}\int_{0}^{\pi}
\cos\left({a}/{\sin\beta}\right)\dd\beta\nonumber\\ &= 1-\tfrac12\pi
a\left[J_0(a) \mathcal{H}_{-1}(a)- J_{-1}(a) \mathcal{H}_0(a)\right],
\end{align}
where $J_n(x)$ and $\mathcal{H}_n(x)$ are the Bessel and Struve functions \citep[2.5.8.6 in][]{PBM91}. 
Finite length of the magnetic structures can be accounted for, but this represents
a small correction and the integral cannot be taken analytically.

To derive $\langle\cos\theta\rangle$ in the simulations,
the particle trajectories were sampled each local Larmor time; the sampling frequency
does not affect the results much
\citep[cf.][]{CH05,RFMB13}. $\langle\cos\theta\rangle$ computed using Eq.~\eqref{ct} and the same obtained from the
simulations show reasonable qualitative agreement if we adopt $d=l_0\Rm^{-1/2}$ for the flow~\eqref{willis_vel}
and $d$ as the thickness of the
magnetic structures calculated using the Minkowski functionals \citep{Wilkin_et_al2007} for the flow~\eqref{KS}. 

Figure~\ref{crw}b,c shows the variation of $\kappa$ with $\langle\cos\theta\rangle$, where $\kappa$ 
is obtained numerically for both the intermittent and randomized magnetic fields, and in each case
the corresponding $\kappa$ predicted from Eq.~\eqref{dcrw} is also shown.
For $\tau$ in Eq.~\eqref{dcrw}, we have used $\RL/v_0$,
where $\RL$ is the local Larmor radius.  The agreement is remarkably good for
the flow \eqref{willis_vel} and excellent for the less regular magnetic field
resulting from the flow \eqref{KS}. This confirms directly that the \CR\
propagation is a CRW with the diffusivity given by
Eq.~\eqref{dcrw}.  This applies to \textit{both} intermittent and
Gaussian random magnetic fields (see also Fig.~\ref{dw}b).  We note that 
the first term in Eq.~\eqref{dcrw} dominates at large $\Lar$.
\section*{Conclusions} We have demonstrated that \CR\ propagation in random
magnetic fields is affected by magnetic intermittency in the range of
energies \eqref{range}, or  \[ \frac{E}{1\GeV} \lesssim 10^9\frac{l_0}{1\kpc}\,
\frac{B}{1\mkG}.  \] In the interstellar medium, $l_0\simeq100\p$ and
$B\simeq10\mkG$ and for ultra-relativistic protons, this energy range is $E\lesssim10^{9}\GeV$. 
In galaxy clusters,
$l_0\simeq10\kpc$, $B\simeq2\mkG$ and $E\lesssim10^{10}\GeV$.

Assuming $R_{\rm m,eff} = \Rmc = 100$ in the interstellar medium, we might expect some effect at
$\RL/l_0=0.1$, which would correspond to $10^8\GeV$ protons using the above
values. Such an effect might produce a knee or spectral break in the cosmic ray
energy spectrum near this energy. The influence of magnetic intermittency
extends to below this energy (the effect of intermittency on cosmic ray diffusivity increases as energy decreases), 
but further investigation is needed to quantify this.
Finally, we note that magnetic intermittency may also affect ultrahigh energy cosmic rays that 
propagate non-diffusively, and that their propagation can also be interpreted as a CRW.
\begin{acknowledgments} We are grateful to J.\ Rachen, A.\ van Vliet and
L.~F.~S.~Rodrigues for useful discussions. This work was
supported by the Thailand Research Fund (RTA5980003), the Leverhulme Trust (RPG-2014-427) and STFC
(ST/N000900/1, Project 2).
\end{acknowledgments}


\end{document}